\documentclass{article}
\usepackage{ijcai26}

\usepackage{times}
\usepackage{soul}
\usepackage{url}
\usepackage[hidelinks]{hyperref}
\usepackage[utf8]{inputenc}
\usepackage[small]{caption}
\usepackage{graphicx}
\usepackage{amsmath}
\usepackage{amsthm}
\usepackage{amssymb}
\usepackage{booktabs}
\usepackage{algorithm}
\usepackage{algorithmic}
\usepackage[switch]{lineno}
\usepackage{multirow}

\title{MP2D: Constrained Monte Carlo Tree-Guided Diffusion for Multi-Objective Protein Sequence Design}

\author{
Zitai Kong$^{1,\dagger}$\and
Yifan Dong$^{2,\dagger}$\and
Yixuan Wu$^3$\and
Zhaokang Liang$^1$\and
Jian Wu$^{1,3,4,*}$\And
Hongxia Xu$^{3,*}$\\
\affiliations
$^1$College of Computer Science and Technology, Zhejiang University, Hangzhou, China\\
$^2$School of Mathematical Sciences, Zhejiang University, Hangzhou, China\\
$^3$State Key Laboratory of Transvascular Implantation Devices and TIDRI, Hangzhou, China\\
$^4$Zhejiang Key Laboratory of Medical Imaging Artificial Intelligence, Hangzhou, China\\
\emails
\{kongzitai, 3230101152, wyx\_chloe, zhaokangliang, wujian2000, einstein\}@zju.edu.cn
}

\begin{document}

\maketitle

\begin{abstract}
    Designing functional protein sequences that satisfy multiple desired properties is a core research focus of protein engineering. Prior methods struggle with inability or inefficiency when dealing with numerous, often conflicting, properties. We propose \textbf{Multi-Property Protein Diffusion, (MP2D)}, a unified framework for multi-objective protein sequence optimization that integrates conditional discrete diffusion with constrained MCTS and global iterative refinement. MP2D formulates diffusion denoising as a constrained sequential decision-making process and employs MCTS to explore diverse denoising trajectories guided by Pareto-based rewards. A global iterative refinement strategy further enables repeated remasking and re-optimization of candidate sequences, while a dynamic Pareto constraint prevents candidate bloat and maintains balanced trade-offs across objectives. We evaluate MP2D on two challenging multi-objective protein design tasks: antimicrobial peptide and protein binder optimization, involving four to five conflicting properties. Experimental results demonstrate that MP2D consistently outperforms existing multi-objective baselines, achieving robust and balanced improvements across all objectives without retraining generative models. These results highlight MP2D as a practical and scalable solution for multi-objective functional protein design.
\end{abstract}

\begingroup
\renewcommand{\thefootnote}{}\footnotetext{\\$^\dagger$ Contributed equally
\\$^*$ Corresponding authors
}
\endgroup

\begin{figure*}[ht]
\centering
\includegraphics[width=0.9\textwidth]{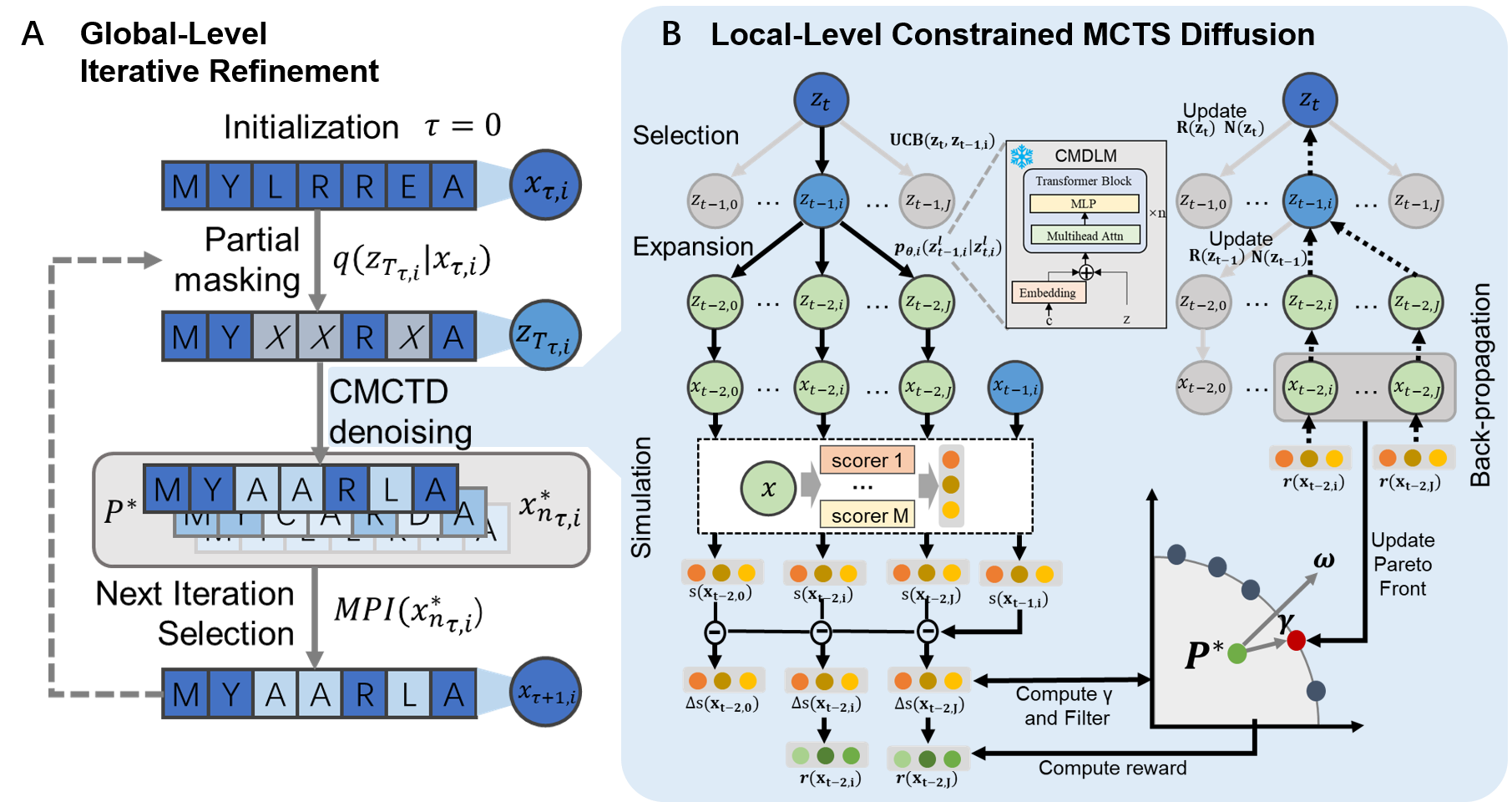} 
\caption{Overview of MP2D. (A) Illustration of global-level iterative refinement process. (B) Visualization of the constrained MCTS-guided diffusion process and the CMDLM model structure.}
\label{fig1}
\end{figure*}

\section{Introduction}

The design of functional protein sequences often requires the simultaneous optimization of multiple properties, such as activity, stability, toxicity, and specificity. In many real-world applications, including antimicrobial peptides (AMPs)~\cite{wang2025hmamp} and therapeutic protein binders (PBs)~\cite{fleishman2011computational}, these properties are inherently conflicting~\cite{tokuriki2009protein}, making multi-objective optimization a fundamental challenge in protein engineering. Consequently, developing computational methods that can efficiently balance multiple protein properties is critical for practical protein design~\cite{nanda2017searching}.

Traditional multi-objective optimization approaches typically extend single-property optimization by applying static weighting schemes or hierarchical optimization strategies~\cite{liu2025monte,yu2023multi}. However, such methods are highly sensitive to weight selection and often fail to maintain balanced trade-offs among conflicting objectives~\cite{chen2023weights,lu2025learning}. Pareto-based optimization offers a more principled alternative by explicitly modeling trade-offs among objectives. Nevertheless, as optimization progresses, the Pareto frontier can rapidly bloat and sparsify, which significantly hinders efficient search and makes it difficult to identify high-quality solutions—especially when optimizing more than three or strongly conflicting properties~\cite{giovanelli2024interactive}. As a result, existing methods struggle to scale to realistic multi-objective protein design problems.

Recent advances in generative models have substantially improved the efficiency of protein sequence design~\cite{alamdari2023protein}. To further guide generation toward desirable outcomes, several studies integrate planning algorithms, such as Monte Carlo Tree Search (MCTS), into the inference process of generative models~\cite{liu2025monte}. These training-free approaches enable flexible optimization with plug-and-play reward functions. However, they heavily rely on the initial quality of the generative model and typically perform only single-shot generation, which limits their ability to correct early suboptimal decisions and explore diverse trade-offs~\cite{zhao2024efficient,jensen2019graph}.

A key observation is that multi-objective protein design inherently admits multiple valid optimization paths, while global property evaluations are often noisy and imperfect. This motivates a search-based and iterative optimization paradigm that can both explore diverse trade-offs and progressively refine candidate solutions. Moreover, effective multi-objective optimization requires explicit mechanisms to control the growth and balance of Pareto candidates.

Here, we propose \textbf{Multi-Property Protein Diffusion (MP2D)}, a training-free framework for multi-objective protein sequence optimization. MP2D combines a classifier-free conditional discrete masked diffusion language model for sequence generation with a constrained MCTS-based diffusion refinement process at inference time. A global iterative refinement strategy repeatedly reselects, partially remasks, and re-optimizes candidate sequences, enabling continuous correction of earlier decisions. In addition, MP2D integrates a dynamic constraint on Pareto frontier updates to prevent candidate bloat and maintain balanced optimization across objectives. Our contributions are summarized as follows:

\begin{itemize}
\item \textbf{A unified design framework for multi-objective protein sequence optimization}. We propose MP2D, a unified framework that balances multiple conflicting protein properties without retraining generative models through constrained MCTS-guided conditional discrete diffusion with global iterative refinement.
\item \textbf{Conditional masked diffusion language model for protein sequence generation (CMDLM)}. We introduce a classifier-free label-guided conditional masked diffusion language model that provides high-quality, task-specific protein sequence generation as a foundation for optimization.
\item \textbf{Iterable training-free multi-objective optimization with MCTS}. We propose an inference-time MCTS-guided diffusion framework that supports iterative refinement through global resampling and remasking, enabling robust exploration of multi-objective trade-offs without retraining.
\item \textbf{Dynamic Pareto frontier constraint for MCTS diffusion (CMCTD)}. We develop a strategy to constrain Pareto frontier updates during MCTS diffusion, preventing optimization collapse caused by frontier bloat.
\item \textbf{Remarkable empirical performance on functional protein design}.  We demonstrate that MP2D effectively balances four to five conflicting properties on antimicrobial peptide and protein binder design tasks, consistently outperforming existing multi-objective baselines.
\end{itemize}

\section{Preliminaries}

This section provides a brief overview of the key concepts underlying our approach, including discrete masked diffusion language models, Pareto-based multi-objective optimization, and Monte Carlo Tree Search (MCTS). These preliminaries introduce the notation and standard formulations required for the subsequent sections.

\subsection{Discrete Masked Diffusion Language Models}

Let $Cat(x; p)$ denote a categorical distribution over a discrete sequence $x=[x_0, x_1, \dots, x_L]$, parameterized by a probability simplex vector $p\in\mathbf{R}^{|\mathcal{V}|}$. The unconditional masked diffusion framework for discrete sequence generation~\cite{ho2020denoising} defines a forward noising process $q(x^{(t)}|x^{(t-1)})$ as a Markov process for $t=0, \dots, T$, which progressively corrupts the input sequence. The reverse denosing process $p_\theta(x^{(t-1)}|x^{(t)})$ is parameterized by a neural network with parameters $\theta$, and aims to recover less corrupted sequences.
\begin{equation}
    \resizebox{.85\linewidth}{!}{$
            \displaystyle
            q(x^{(t)}|x^{(t-1)}) = Cat(x^{(t)}; \beta_tx^{(t-1)}+(1-\beta_t)q_{nosie}(x^{(t)})
        $}
    \label{eqn1}
\end{equation}
where $q_{nosie}$ denotes a stationary noise distribution over the vocabulary, and $0\ll\beta_t<1$ is  a noise schedule controlling the corruption level at diffusion step $t$. The marginal distribution from the original sequence $x^{(0)}$ admits a closed-form expression with $\alpha_t=\prod^{t}_{i=1}\beta_i$:
\begin{equation}
    q(x^{(t)}|x^{(0)}) = Cat(x^{(t)}; \alpha_tx^{(0)}+(1-\alpha_t)q_{nosie}(x^{(t)})
    \label{eqn2}
\end{equation}

During inference, new sequences are generated by iteratively applying the learned reverse process
\begin{equation}
    p_\theta(x^{(t-1)}|x^{(t)}) = \sum_{\hat{x}_0}q(x^{(t-1)}|x^{(t)}, \hat{x}_0)p_\theta(\hat{x}_0|x^{(t)})
    \label{eqn3}
\end{equation}
where $\hat{x}_0$ is first sampled from $p_\theta(\cdot|x^{(t)})$ and a less noisy $x^{(t-1)}$ is subsequently sampled by $q(\cdot|x^{(t)}, x^{(0)}=\hat{x}_0)$.

\subsection{Pareto Optimization}

Optimizing multiple objectives, especially conflicting properties in functional protein design, typically involves severe trade-offs. A widely adopted solution is to identify a set of sequences that cannot be further improved in any single objective without degrading performance in at least one other objective; such sequences are referred to as \textbf{Pareto-optimal}~\cite{ngatchou2005pareto}. Suppose each sequence $x$ is evaluated on $M$ objectives with corresponding score functions, denoted by a vector $s(x) = [s_1(x), \dots, s_M(x)]\in \mathcal{R}^M$, A sequence $x^*$ is said to dominate another sequence $x$ as
\begin{equation}
\resizebox{.9\linewidth}{!}{$
    \begin{split}
        s(x^*)\succ s(x) &\text{  iff  } \forall m\in[1,\dots,M] \text{ s.t. } s_m(x^*)\geq s_m(x) \\
        & \land \exists m' \in [1,\dots,M] \text{ s.t. } s_{m'}(x^*)> s_{m'}(x)
    \end{split}
    $}
    \label{eqn4}
\end{equation}
The set of all sequences that are not dominated by any other sequence is called the \textbf{Pareto front}, defined as
\begin{equation}
    \mathcal{P}^* = \{x|\nexists x^*\in \mathcal{P}^* \text{ s.t. } s(x^*)\succ s(x)\}
    \label{eqn5}
\end{equation}

\subsection{Monte Carlo Tree Search}

MCTS~\cite{chaslot2008monte} is a planning algorithm that combines tree search with stochastic simulation. It balances exploration and exploitation in the decision space and iteratively refines the search toward high-reward regions. A standard MCTS procedure consists of four stages: selection, expansion, simulation, and backpropagation. Starting from the root node, the selection stage recursively chooses child nodes according to the Upper Confidence Bound (UCB) criterion
\begin{equation}
    \text{UCB} = Q(s,a)+c\sqrt{\frac{\log N(s)}{N(s,a)}}
    \label{eqn6}
\end{equation}
where $Q$ denotes the estimated reward, $c$ controls the exploration strength, and $N$ represents the visit counts, respectively. In the expansion stage, new child nodes corresponding to previously unvisited actions are added to the tree. The simulation stage performs rollouts from the expanded node to a terminal state to obtain an evaluation. The backpropagation stage updates the $Q$ and $N$ values along the visited path.

\section{Methods}

In this section, we propose a training-free framework for multi-objective protein sequence optimization based on conditional discrete diffusion, shown in Figure 1. We cast diffusion denoising as a sequential decision-making problem and guided by a dynamically constrained MCTS with Pareto-based rewards. We design a global iterative refinement, which further improves robustness by repeatedly masking and re-optimizing candidate sequences. The implementation details and hyperparameter settings can be found in Appendix A.

\subsection{Conditional masked diffusion language model}

Directly applying pretrained diffusion language models trained on large-scale, general protein corpora often results in an excessively large search space, which makes subsequent reward-guided generation inefficient and may hinder exploration of valid local protein regions. To address this issue, we extend the unconditional masked diffusion language model for discrete sequences to CMDLM, a label-guided, classifier-free conditional diffusion model tailored to specific functional protein design tasks. We introduce a condition label $c$, representing the protein type, into the diffusion model and parameterize the reverse process as $p_\theta(x^{(0)}|x^{(t)}, c)$. To enable classifier-free guidance, the model combines conditional and unconditional predictions~\cite{ho2022classifier} with
\begin{equation}
    \hat{\textit{l}} = (1+w)p_\theta(x^{(0)}|x^{(t)},c) - wp_\theta(x^{(0)}|x^{(t)},\emptyset)
    \label{eqn7}
\end{equation}
where $w\geq0$ controls the guidance strength and $\emptyset$ denotes the unconditional condition. The resulting guided distribution is
\begin{equation}
    \tilde{p}_\theta(x^{(0)}|x^{(t)},c) = softmax(\hat{\textit{l}}) \propto\frac{p_\theta(x^{(0)}|x^{(t)},c)^{1+w}}{p_\theta(x^{(0)}|x^{(t)},\emptyset)^w}
    \label{eqn8}
\end{equation}

In practice, we jointly train the conditional and unconditional models by randomly replacing $c$ with the unconditional identifier $\emptyset$ with probability $p_{uncond}$. The training objective is a reweighted cross-entropy loss
\begin{equation}
    \mathcal{L}_t = \mathbb{E}_{q(x^{(0)})}[\lambda^{(t)}\sum^{L}_{i=1}b_i(t)\log{\tilde{p}_\theta(x_i^{(0)}|x^{(t)},c)}]
    \label{eqn9}
\end{equation}
where $\lambda^{(t)}$ is a timestep-dependent weight induced by the noise schedule, and $b_i(t)=\mathbb{I}[x_i^{(t)}\neq x_i^{(0)}]$ indicates corrupted positions. During generation, the initial sequence is fully masked. At each diffusion step, a subset of masked tokens is updated by sampling
\begin{equation}
    p_\theta(x^{(t-1)}|x^{(t)}) = \sum_{\hat{x}_0}q(x^{(t-1)}|x^{(t)}, \hat{x}_0)\tilde{p}_\theta(\hat{x}_0|x^{(t)},c)
    \label{eqn10}
\end{equation}
This conditional model serves as the backbone for our subsequent constrained MCTS-based diffusion framework.

\subsection{Constrained MCTS Diffusion}
Given the conditional diffusion backbone, we formulate the denoising process as a sequential decision-making problem and employ MCTS to explore multiple denoising trajectories under flexible multi-objective constraints. The proposed framework relies exclusively on pretrained models and external property evaluators, requiring no additional training while enabling scalable optimization over an arbitrary number of objectives through constraint-based filtering. By integrating global objective evaluation with local diffusion transitions, this module supports efficient exploration of the constrained protein design space. Detailed algorithm is in Appendix C.4

\noindent \textbf{Initialization} Let $z_t$ denote a partially unmasked sequence at diffusion step $t$, and let $child(z_t) = \{z_{t-1,0}, \dots, z_{t-1,J}\}$ denote its child nodes generated by one-step denoising. To enable global iterative refinement from candidate proteins, we initialize the MCTS root as a partially masked sequence $z_T\in |\mathcal{V}|^L$ with noise level $T$. The Pareto set $\mathcal{P^*}$  is initialized as empty. We assume $M$ pretrained or selected scoring functions $\{s_m(\cdot)\}^M_{m=1}$. To control optimization trade-offs, we construct a set of direction vectors $\omega \in \mathbf{R}^M$ using the Das–Dennis simplex lattice with $H$ subdivisions:
\begin{equation}
    \omega_i = \frac{q_i}{H}, \text{ where } q_i \in \textbf{Z}_{\geq0}, \sum^{M}_{i=1}q_i = H
    \label{eqn11}
\end{equation}

\noindent \textbf{Selection} Starting from the root, the tree is traversed until a leaf node is reached. At each intermediate node $z_t$, we restrict candidate actions to its Pareto non-dominated children:
\begin{align}
    \begin{split}
        \mathcal{P}^*_{z_t} = & \{z_{t-1,i}|\nexists z_{t-1,j} \in \text{child}(z_t)\\ & \text{ s.t. } UCB(z_t, z_{t-1,j})\succ UCB(z_t, z_{t-1,i})\}
    \end{split}
    \label{eqn12}
\end{align}
One child is uniformly sampled from $\mathcal{P}^*_{z_t}$ and the modified UCB score is defined as
\begin{equation}
\resizebox{.95\linewidth}{!}{$
    UCB(z_t, z_{t-1,i}) = \frac{R(z_{t-1,i})}{N(z_{t-1,i})}+cp_\theta(z_{t-1,i}|z_t)\frac{\sqrt{N(z_t)}}{1+N(z_{t-1,i})}
$}
\label{eqn13}
\end{equation}
where $R(\cdot)$ and $N(\cdot)$ denote the cumulative reward vector and visit count, respectively. The first term promotes exploitation of high-reward nodes, while the second term encourages exploration guided by the diffusion posterior to ensure validity.

\noindent \textbf{Expansion} For a selected leaf node $z_t$ that is not fully unmasked, we generate $J$ child nodes $z_{t-1}$ by sampling from the diffusion model $p_\theta(z_{t-1,i}|z_t)$. To avoid duplicate generations, we inject independent Gumbel noise into the token-level logits~\cite{tang2025peptune}:
\begin{equation}
    p_{\theta,i}(z_{t-1,i}^l|z_t^l) = \log p_{\theta}(z_{t-1,i}^l|z_t^l) + \mathcal{G}_i^l
    \label{eqn14}
\end{equation}
where $\mathcal{G}_i^l = -\log(-log(u_{i,j}^l+\epsilon)+\epsilon)$, $u_{i,j}^l\sim \text{Uniform}(0,1)$.

\noindent \textbf{Simulation and Constraint Filtering} Each child node $z_{t-1,i}$ is fully denoised via greedy decoding to obtain a clean sequence $x_{t-1,i}$, which is evaluated by all scoring functions to yield $s(x_{t-1, i})=[s_1(x_{t-1, i}), \dots, s_M(x_{t-1, i})]$.

To prevent rapid expansion of the search tree and uncontrolled growth of the Pareto set, we constrain candidate additions using predefined optimization directions. We compute the improvement vector
\begin{equation}
    \Delta s_m(x_{t-1,i}) = s_m(x_{t-1,i}) - s_m(x_t)
    \label{eqn15}
\end{equation}
and its cosine similarity with direction $\omega$:
\begin{equation}
    \gamma(z_{t-1,i}) = \frac{\Delta s(x_{t-1,i})\cdot \omega}{\left\| \Delta s(x_{t-1,i}) \right\|\left\| \omega \right\|}
    \label{eqn16}
\end{equation}
Only candidates satisfying $arccos(\gamma) \leq \Psi$ are retained; fallback rules are applied when no such candidates exist (see Appendix C.1). To avoid excessive rejection or overly permissive filtering during different stages of the search, the angular threshold $\Psi$ can be adaptively adjusted to maintain a stable acceptance rate (see Appendix C.2).

For each retained sequence, the reward vector $r(x_{t-1, i})=[r_1(x_{t-1, i}), \dots, r_M(x_{t-1, i})]\in \textbf{R}^M$ is computed as
\begin{equation}
    r_m(x_{t-1,i}) = \frac{1}{|\mathcal{P}^*|}\sum^{|\mathcal{P}^*|}_{n=1} \mathbf{1}[s_m(x_{t-1,i}) \geq s_m(x_n^*)]
    \label{eqn17}
\end{equation}
indicating its relative dominance over current Pareto candidates. Finally, the Pareto set is updated by adding non-dominated sequences and removing dominated ones
\begin{equation}
    \mathcal{P}^* = \mathcal{P}^* \cup \{x_{t-1,i}|\forall x_n^*\in\mathcal{P}^* \text{ } s(x_{t-1,i}) \succeq s(x_n^*)\}
    \label{eqn18}
\end{equation}
\begin{equation}
    \mathcal{P}^* = \mathcal{P}^* / \{x_n^*|\exists x_{t-1,i} \text{ s.t. } s(x_{t-1,i}) \succeq s(x_n^*)\}
    \label{eqn19}
\end{equation}

\noindent \textbf{Back-propagation} For a newly explored node $z_{t-1,i}$, we update the cumulative reward vector as $R(z_{t-1,i}) = r(x_{t-1,i})$ and its visit count as $N(z_{t-1,i}) = 1$. For each ancestor node $z_t$ along the path to the root, we update
\begin{equation}
    R(z_t) = R(z_t) + \sum_{i=1}^J r(x_{t-1,i})
    \label{eqn20}
\end{equation}
\begin{equation}
    N(z_t) = N(z_t) + 1
    \label{eqn21}
\end{equation}

\begin{table*}[htb]
\centering
\caption{Performance comparison between CMDLM and baseline models on three tasks: Uniprot peptides, protein binders and AMPs.}
\resizebox{0.9\linewidth}{!}{
\renewcommand\arraystretch{1.1}
\begin{tabular}{l|ccc|ccc|ccc}
\hline
\multirow{2}{*}{Methods}        & \multicolumn{3}{c|}{Uniprot Peptide} & \multicolumn{3}{c|}{Protein Binder} & \multicolumn{3}{c}{AMP} \\ 
                                & pLDDT($\uparrow$)     & pppl($\downarrow$)    & FPD($\downarrow$)    & pLDDT($\uparrow$)       & pppl($\downarrow$)       & FPD($\downarrow$)      & pLDDT($\uparrow$)   & pppl($\downarrow$)   & FPD($\downarrow$)  \\ \hline
{dataset}    & 71.35         & 12.57       & 0.2317      & 71.45           & 12.58          & 0.1961        & 73.69       & 12.05      & 0.2617    \\ \hline
{ProteinGAN} & 67.84         & 15.13       & 1.5067      & 65.15           & 15.90          & 2.3336        & 60.23       & 14.37      & 2.7669    \\
{ProtGPT2}   & 70.86         & 12.68       & 3.3646      & 70.14           & 13.39          & 3.1198        & 70.61       & \textbf{13.19}      & 3.5444    \\
{EvoDiff}    & 64.99         & 14.25       & 0.6011      & 68.96           & 14.39          & 0.5944        & 69.68       & 13.51      & 0.6352    \\ \hline
{CMDLM}      & \textbf{73.12}         & \textbf{11.44}       & \textbf{0.3280}      & \textbf{70.68}           & \textbf{13.32}          & \textbf{0.4106}        & \textbf{71.84}       & 13.33      & \textbf{0.5423}    \\ \hline
\end{tabular}
}
\label{table1}
\end{table*}

\subsection{Global-Level Iterative Refinement}

Although a single-shot CMCTD step can produce improved sequences, it still faces two limitations. First, global property evaluators may introduce noisy or inconsistent assessments, and once a token is unmasked, the decision cannot be revised within the same denoising trajectory. This makes single-pass optimization vulnerable to early errors. Second, different masking patterns can lead to different optimization paths, and relying on a single denoising trajectory reduces candidate diversity and may miss better solutions. Detailed algorithm can be found in Appendix C.3.

To address these issues, we introduce a global iterative refinement framework, where sequences are progressively improved through multiple cycles of partial masking and constrained MCTS-based denoising. This enables more robust optimization of global properties and allows continual correction of earlier decisions.

\noindent \textbf{Initialization} Let the refinement process start with $W$ seed protein sequences $\{x_{\tau,1}, \dots, x_{\tau,W}\} \in \mathcal{V}^{W\times L}$, with $\tau=0$, where $\tau \in [0,\tau_{max}]$ denotes the refinement iteration. At each iteration, a noise level $T_{\tau,i}$ is sampled for each sequence.

\noindent \textbf{Partial Masking} For every sequence $x_{\tau,i}$, we construct a partially masked sequence $z_{T_{\tau,i}}$ with
\begin{equation}
\resizebox{.95\linewidth}{!}{$
    q(z_{T_{\tau,i}}|x_{\tau,i}) = Cat(z_{T_{\tau,i}}; \alpha_{T_{\tau,i}}x_{\tau,i}+(1-\alpha_{T_{\tau,i}})q_{noise}(z_{T_{\tau,i}})
$}
    \label{eqn22}
\end{equation}
where $\alpha_{T_{\tau,i}}=\prod^{T_{\tau,i}}_{i=1}\beta_i$. This masked sequence serves as the starting point for constrained MCTS diffusion.

\noindent \textbf{MCTS-Based Denoising} A single constrained MCTS diffusion step is applied to each partially masked sequence, producing a set of Pareto non-dominated candidates $\mathcal{P}^*$.

\noindent \textbf{Next Iteration Selection} Despite the constraint mechanism, the Pareto set may still grow large ($|\mathcal{P}^*|>W$). To select a representative next-iteration seed, we evaluate each candidate $x^*_{n,\tau,i}\in \mathcal{P}^*$ by computing the improvement vector $\Delta s(x^*_{n,\tau,i})$ with
\begin{equation}
    \Delta s_m(x^*_{n,\tau,i}) = s_m(x^*_{n,\tau,i}) - s_m(x_{\tau,i})
    \label{eqn23}
\end{equation}

We then define a multi-property improvement (MPI) score
\begin{equation}
    \begin{split}
        MPI(x^*_{n,\tau,i}) & = \text{z-score}(\frac{1}{M}\sum^M_{m=1}\mu_m\text{Rank}_m(x^*_{n,\tau,i})) \\
        &+ \lambda\text{z-score}(D(x^*_{n,\tau,i}))
    \end{split}
    \label{eqn24}
\end{equation}
where
\begin{equation}
    \text{Rank}_m(x^*_{n,\tau,i})=\frac{rank(\Delta s_m(x^*_{n,\tau,i}))}{T_{\tau,i}}
    \label{eqn25}
\end{equation}
\begin{equation}
    D(x^*_{n,\tau,i})=\Delta s(x^*_{n,\tau,i}) \cdot \omega
    \label{eqn26}
\end{equation}
Here, $\mu$ is normalized weights ($\sum^M_{m=1}\mu_m=1$), $\text{Rank}_m(\cdot)$ measures the relative improvement on each objective, where the $T_{\tau,i}$ division is to ensure comparability across different corruption levels. $D(\cdot)$ measures alignment with the predefined optimization direction. Finally, the next-iteration sequence is sampled from the Pareto set according to:
\begin{equation}
    x_{\tau+1,i}\sim \frac{\exp{MPI(x^*_{n,\tau,i})}}{\sum_{x\in\mathcal{P}^*}\exp{MPI(x)}}
    \label{eqn27}
\end{equation}

\section{Experiments}

We pretrained CMDLM on general peptides and performed conditional finetunes and multi-property optimizations on two important protein medicines, protein binders (PBs) and antimicrobial peptides (AMPs). In this section, we first discuss the benchmarks used. Then we introduce and analysis the performance of the CMDLM generation and the CMCTD optimization on both cases. Additional visualization analysis can be found in Appendix D.

\subsection{Experimental Setup} 

\subsubsection{Datasets}

For pretraining CMDLM, we collected peptide sequences of length 2–50 from UniProt~\cite{uniprot2019uniprot}, yielding 2.6M sequences after removing those containing uncommon amino acids. 
For antimicrobial peptide (AMP) design, we followed the data collection procedure in~\cite{chen2024amp} and combined sequences from dbAMP~\cite{yao2025dbamp}, AMP Scanner~\cite{veltri2018deep}, and DRAMP~\cite{shi2022dramp}. After filtering for length ($<40$) and standard amino acids, we obtained 195k peptides.
For protein binder (PB) design, we adopted the dataset from~\cite{chen2025multi}, which includes 15.5k peptides curated from PepNN, BioLip2, and PPIRef. Sequences range from 6 to 49 residues, and those with uncommon amino acids were removed. All datasets are splited into training and validation sets with a ratio of 9:1.

\subsubsection{Baselines}

To evaluate CMDLM, we followed the benchmark protocol of~\cite{meshchaninov2024diffusion}, comparing against: ProteinGAN~\cite{repecka2021expanding} (GAN-based), EvoDiff~\cite{alamdari2023protein} (discrete diffusion), ProtGPT2~\cite{ferruz2022protgpt2} (autoregressive). All baseline models were configured with comparable parameter sizes and standard character-level tokenization to ensure fairness.

For the PB optimization task, we followed the benchmark of~\cite{chen2025multi}, including four classical multi-objective evolutionary algorithms—NSGA-III, SMS-EMOA, SPEA2, and MOPSO—and a recent discrete flow matching–based method MOG-DFM, specifically introduced for protein binder optimization.

To the best of our knowledge, no established benchmark exists for multi-objective AMP optimization. We therefore constructed one using advanced AMP design models that are reported to be able to optimize over 3 properties, including Multi-CGAN~\cite{yu2023multi}, MPOGAN~\cite{liu2025multi}, HMAMP~\cite{wang2025hmamp}, and the multimodal fusion model MoFormer~\cite{wang2024moformer}.

Detailed descriptions and selection reasons of baseline methods can be found in Appendix E.

\subsubsection{Metrics} 

\textbf{Foundation Model Evaluation} We used three metrics to assess CMDLM's performance: \textbf{ESM-2 perplexity}~\cite{lin2022language} for sequence plausibility, \textbf{pLDDT}~\cite{jumper2021highly} for structural foldability, and \textbf{Frechet ProtT5 Distance (FPD)} for distributional similarity.

\textbf{Protein Binder Optimization} We evaluated five therapeutically relevant properties: \textbf{hemolysis}, \textbf{non-fouling}, \textbf{solubility}, \textbf{half-life}, and \textbf{binding affinity}. Predictions were made using the public classifiers provided in~\cite{liu2025multi}.

\textbf{AMP Optimization} We used publicly available predictors to evaluate key properties: AMP-Scanner~\cite{veltri2018deep} for \textbf{antimicrobial probability}, HemoPI~\cite{chaudhary2016web} for \textbf{hemolysis}, and ToxinPred2~\cite{rathore2024toxinpred} for \textbf{toxicity}. we also evaluated \textbf{MIC values} using a regressor trained on MIC data collected from GRAMPA~\cite{plisson2022overcoming} and DBAASP~\cite{pirtskhalava2021dbaasp} (see Appendix B). In addition, detailed definitions and selection criteria can be found in Appendix F.

\subsection{Evaluation of Conditional Diffusion Backbone}

As the generative backbone for subsequent optimization, CMDLM is expected to capture class-specific sequence patterns while producing valid and realistic protein sequences. Due to the limited availability of labeled AMP and protein binder data, we first pretrain the model unconditionally on large-scale peptide sequences to learn general peptide syntax, and then fine-tune it with LoRA under conditional supervision to specialize in target protein classes.

We evaluate CMDLM from three complementary aspects: sequence plausibility, structural foldability, and distributional alignment. Sequence quality is assessed using ESM‑2 perplexity, which measures consistency with natural protein statistics. Structural plausibility is evaluated by the predicted local distance difference test (pLDDT) score. Distributional similarity is measured via the Frechet ProtT5 Distance (FPD) between generated sequences and real proteins from the corresponding classes. Across all evaluation settings, CMDLM almost consistently achieves lower ESM‑2 perplexity and FPD, together with higher pLDDT, compared to baseline methods. These results indicate that CMDLM generates sequences that are statistically reliable, structurally foldable, and well aligned with natural protein distributions, making it a strong foundation for downstream optimization through diffusion-based denoising.

\subsection{Evaluation of Multi-objective Optimization}

\subsubsection{MP2D Generates PBs Satisfying Multiple Properties}

Following the benchmark of~\cite{liu2025multi}, we designed peptide binders for two target proteins: 1B8Q, a smaller protein with a known binder, and PPP5, a larger protein without characterized binders. We first sampled 100 binder candidates of random lengths from CMDLM conditioned on the PB label, and optimized five key therapeutic properties—hemolysis, non‑fouling, solubility, half‑life, and binding affinity—for 100 refinement iterations. Each baseline method generated the same number of candidates per target for a fair comparison. As shown in Table 2, MP2D achieves the lowest hemolysis, the highest non‑fouling, solubility, half‑life, and competitive binding affinity, consistently outperforming all baselines. Figure S2 shows that MP2D-designed binders surpass natural ones across all properties.

Importantly, classical baseline methods are unable to optimize all five conflicting properties simultaneously, often collapsing one or more properties when improving others. In contrast, MP2D remains stable across all objectives, demonstrating its ability to navigate complex trade-offs and maintain balanced improvements.

\begin{table}[t]
\caption[Performance on protein binder design of MP2D and baseline methods]{Performance on protein binder design of MP2D and baseline methods. $\dagger$: benchmark results quoted from~\cite{liu2025multi}.}
\centering
\resizebox{\linewidth}{!}{
\renewcommand\arraystretch{1.7}
\begin{tabular}{ccccccc}
\hline
Target                & Methods  & Hemolysis($\downarrow$)       & Non-Fouling($\uparrow$)     & Solubility($\uparrow$)      & Half-Life($\uparrow$)        & Affinity($\uparrow$)        \\ \hline
\multirow{6}{*}{1B8Q} & MOPSO$\dagger$    & 0.1066          & 0.4763          & 0.4684          & 4.449            & 6.0594          \\
                      & NSGA-III$\dagger$ & 0.0862          & 0.5715          & 0.5825          & 7.324            & 7.2178          \\
                      & SMS-EMOA$\dagger$ & 0.1196          & 0.3450          & 0.3511          & 3.023            & 5.9550          \\
                      & SPEA2$\dagger$    & 0.0819          & 0.4973          & 0.5057          & 4.126            & \textbf{7.3240} \\
                      & MOG-DFM$\dagger$  & 0.0785          & 0.8445          & 0.8455          & 27.227           & 5.9094          \\ \cline{2-7}
                      & MP2D     & \textbf{0.0460} & \textbf{0.8928} & \textbf{0.8977} & \textbf{29.032}  & 6.7794          \\ \hline
\multirow{6}{*}{PPP5} & MOPSO$\dagger$    & 0.0883          & 0.4711          & 0.4255          & 1.769            & 6.6958          \\
                      & NSGA-III$\dagger$ & 0.0479          & 0.7138          & 0.7066          & 2.901            & 7.3789          \\
                      & SMS-EMOA$\dagger$ & 0.1242          & 0.4269          & 0.4334          & 1.031            & 6.2854          \\
                      & SPEA2$\dagger$    & 0.0555          & 0.6221          & 0.6098          & 2.613            & \textbf{7.6253} \\
                      & MOG-DFM$\dagger$  & 0.0617          & 0.7738          & 0.7510          & 27.775           & 6.8197          \\ \cline{2-7}
                      & MP2D     & \textbf{0.0472} & \textbf{0.8900} & \textbf{0.8866} & \textbf{41.5726} & 7.2007          \\ \hline
\end{tabular}
}
\end{table}

\subsubsection{MP2D Designs Optimized AMPs on Scalar Properties}

We performed a parallel procedure for AMP optimization. CMDLM generated 100 candidate peptides under the AMP condition with random lengths, which were then optimized jointly for four critical properties: antimicrobial activity, MIC value, hemolysis, and toxicity. Baseline models were used to generate 100 AMP candidates under the same settings. As summarized in Table 3, MP2D produces AMPs with the highest antimicrobial probability and the lowest MIC, hemolysis and toxicity. Consistently with the binder case, MP2D-designed AMPs also outperform natural AMPs across all four properties (Figure S2). 

Across baselines, we again observe that no existing method can simultaneously improve all four conflicting AMP properties, often leading to performance degradation in at least one objective. MP2D avoids this collapse and produces balanced improvements across all objectives, underscoring its robustness in handling strongly conflicting biological properties.

\begin{table}[h!]
\caption{Performance on AMP of MP2D and baseline methods.}
\centering
\resizebox{\linewidth}{!}{
\renewcommand\arraystretch{1.3}
\begin{tabular}{c|cccc}
\hline
Method     & Pamp($\uparrow$)            & MIC($\downarrow$)             & Hemolysis($\downarrow$)       & Toxicity($\downarrow$)        \\ \hline
Multi-CGAN & 0.3785          & 1.3214          & 0.5813          & 0.7698          \\
MPOGAN     & 0.6201          & 1.2346          & 0.6107          & 0.6166          \\
MoFormer   & 0.4326          & 1.3398          & 0.5761          & 0.8198          \\
HMAMP      & 0.6714          & 1.4266          & 0.5613          & 0.7641          \\ \hline
MP2D       & \textbf{0.9949} & \textbf{0.2689} & \textbf{0.4077} & \textbf{0.2968} \\ \hline
\end{tabular}
}
\end{table}

Moreover, MP2D is training-free and plug-and-play during optimization. By simply replacing the property predictor, MP2D can be directly applied without collecting additional training data or retraining any model—highlighting its practicality and broad applicability.

\subsection{Ablation Studies}

We conduct ablation studies on both protein binder and AMP tasks to examine the contribution of each component in MP2D. Results are summarized in Table 4 and Table 5.

\subsubsection{Conditional Base Model Narrows the Optimization Gap}

We first analyze the impact of initialization strategies and conditional diffusion. Using random peptides as starting points together with the unconditional diffusion model pretrained on general peptides yields the weakest performance across all objectives. Initializing from AMP or protein binder sequences improves optimization quality, indicating that reasonable starting points can partially facilitate search.

In contrast, replacing the unconditional diffusion model with a conditional one leads to substantially larger improvements, even when starting from generic peptides. This demonstrates that conditional diffusion plays a more critical role than initialization alone by constraining the search space to task-relevant protein distributions. Combining AMP or PB initialization with conditional diffusion achieves the best performance, highlighting their complementary effects.

\subsubsection{Constrained Optimization Prevents Property Collapse}

To evaluate the role of constraint filtering, we remove this component while keeping all other modules unchanged. Without constraint filtering, optimization performance degrades noticeably, and several objectives deteriorate despite improvements in others. This indicates that constraint filtering is essential for preventing Pareto front collapse and maintaining balanced trade-offs among conflicting properties.

\subsubsection{Iterative Refinement Gives Further Improvements}

Finally, we assess the effect of global iterative refinement by disabling the remasking-and-reoptimization process. Without iterative refinement, optimization quickly saturates and fails to correct early suboptimal decisions. As a result, both AMP and PB tasks exhibit inferior overall performance compared to the full MP2D framework. These results confirm that global iterative refinement is crucial for robust multi-objective optimization under noisy global evaluations.

\begin{table}[h!]
\caption{Ablation study results on protein binder design task.}
\centering
\setlength{\tabcolsep}{2pt}
\resizebox{\linewidth}{!}{
\renewcommand\arraystretch{1.6}
\begin{tabular}{c|cccc|ccccc}
\hline
Target                & \begin{tabular}[c]{@{}c@{}}start w/\\ PBs\end{tabular} & \begin{tabular}[c]{@{}c@{}}conditional\\ model\end{tabular} & constraint & refined & \begin{tabular}[c]{@{}c@{}}Hemolysis\\ ($\downarrow$)\end{tabular}       & \begin{tabular}[c]{@{}c@{}}Non-Fouling\\ ($\uparrow$)\end{tabular}    & \begin{tabular}[c]{@{}c@{}}Solubility\\ ($\uparrow$)\end{tabular}     & \begin{tabular}[c]{@{}c@{}}Half-Life\\ ($\uparrow$)\end{tabular}       & \begin{tabular}[c]{@{}c@{}}Affinity\\ ($\uparrow$)\end{tabular} \\ \hline
\multirow{5}{*}{1B8Q} &                                                            &             & \checkmark          & \checkmark       & 0.1313    & 0.4660      & 0.4723     & 9.2736    & 5.4521   \\
                      &                                                            & \checkmark           & \checkmark          & \checkmark       & 0.0821    & 0.7174      & 0.7130     & 19.4659   & 5.9339   \\
                      & \checkmark                                                          &             & \checkmark          & \checkmark       & 0.0940    & 0.6442      & 0.6416     & 12.7857   & 5.6422   \\
                      & \checkmark                                                          & \checkmark           &            & \checkmark       & 0.0460    & 0.8928      & 0.8977     & 15.0126   & 6.1130   \\
                      & \checkmark                                                          & \checkmark           & \checkmark          &         & 0.1289    & 0.3377      & 0.3585     & 3.4126    & 5.2340   \\ \hline
\multirow{5}{*}{PPP5} &                                                            &             & \checkmark          & \checkmark       & 0.1290    & 0.4575      & 0.4654     & 8.2453    & 5.9767   \\
                      &                                                            & \checkmark           & \checkmark          & \checkmark       & 0.0823    & 0.7155      & 0.7174     & 18.0461   & 6.4319   \\
                      & \checkmark                                                          &             & \checkmark          & \checkmark       & 0.1089    & 0.5889      & 0.5995     & 7.8438    & 6.0806   \\
                      & \checkmark                                                          & \checkmark           &            & \checkmark       & 0.0590    & 0.8681      & 0.8684     & 27.3747   & 6.7360   \\
                      & \checkmark                                                          & \checkmark           & \checkmark          &         & 0.1138    & 0.4102      & 0.4226     & 2.0140     & 5.0023   \\ \hline
\end{tabular}
}
\end{table}

\begin{table}[h!]
\caption{Ablation study results on the AMP design task.}
\centering
\setlength{\tabcolsep}{3pt}
\resizebox{\linewidth}{!}{
\renewcommand\arraystretch{1.4}
\begin{tabular}{cccc|cccc}
\hline
\begin{tabular}[c]{@{}c@{}}start w/\\ AMPs\end{tabular} & \begin{tabular}[c]{@{}c@{}}conditional\\ model\end{tabular} & constrained               & refined                   & Pamp($\uparrow$)            & MIC($\downarrow$)             & Hemolysis($\downarrow$)       & Toxicity($\downarrow$) \\ \hline
                                                        &                                                             & \checkmark & \checkmark & 0.3302 & 0.9842 & 0.4466    & 0.0064   \\
                                                        & \checkmark                                   & \checkmark & \checkmark & 0.6104 & 0.9035 & 0.4274    & 0.1415   \\
                                                        
\checkmark                               &                                                             & \checkmark & \checkmark & 0.7986 & 0.9526 & 0.4746    & 0.4012   \\
\checkmark                               & \checkmark                                   &                           & \checkmark & 0.9712 & 0.6533 & 0.5400    & 0.3319   \\
\checkmark                               & \checkmark                                   & \checkmark &                           & 0.8489      & 0.7749      & 0.5244         & 0.4774        \\ \hline
\end{tabular}
}
\end{table}

\section{Related Work}

\textbf{Generation models for \textit{de novo} protein sequence design} aim to produce novel protein sequences from scratch. Recent advances have demonstrated promising results using autoregressive models such as ProtGPT2~\cite{ferruz2022protgpt2}, GAN-based approaches such as ProteinGAN~\cite{repecka2021expanding}, and diffusion-based models such as EvoDiff~\cite{alamdari2023protein}. These models are effective at capturing the statistical patterns of natural protein sequences and generating realistic candidates. However, they do not explicitly optimize sequences toward multiple functional objectives during inference. As a result, additional optimization mechanisms are often required to steer generation toward desired property trade-offs.

\textbf{Planning-based generation models} processes toward high-quality outputs under task-specific objectives. MCTS and related lookahead techniques are widely used general mechanism. RethinkMCTS~\cite{li2025rethinkmcts}       applies MCTS to refine reasoning steps in LLMs, while MCTD~\cite{yoon2025monte} integrates diffusion denoising with tree search. In biomolecular design, ProtInvTree~\cite{liu2025protinvtree} and PepTune~\cite{tang2025peptune} similarly combine tree search with denoising processes for inverse folding and molecular design tasks. However, these methods typically guide a single generation trajectory or optimize a limited number of objectives. It can become inefficient due to the combinatorial growth of candidate expansions and the accumulation of weakly non-dominated solutions when extended to numerous objectives. 

\textbf{Multi-objective protein sequence design} seeks sequences that satisfy several desired, and often conflicting, properties. A common strategy is scalarization, which reduces multiple objectives to a single score via weighted sums or hierarchical priorities. MCTD-ME~\cite{liu2025monte} extends MCTD by aggregating multiple properties into a weighted evaluation score. Multi-CGAN~\cite{yu2023multi} conditions generative models on combinations of property labels, while HMAMP~\cite{wang2025hmamp} employs multiple discriminators to enforce property constraints during training. Another paradigm performs model-based sequence optimization using learned predictors. MPOGAN~\cite{liu2025multi} iteratively retrains generators with classifier-filtered samples, and MOG-DFM~\cite{chen2025multi} guides flow matching with adaptive probability paths for multi-objective optimization. While these methods can improve target properties, they typically rely on repeated retraining or careful tuning of optimization signals, and may struggle to maintain balanced trade-offs as the number of objectives increases.

\textbf{Iterative training-free optimization} focuses on progressively refining candidate solutions without retraining underlying models. Classical population-based methods, including swarm-based approaches such as MOPSO~\cite{coello2002mopso} and evolutionary algorithms such as NSGA-III~\cite{ishibuchi2016performance}, SMS-EMOA~\cite{beume2007sms}, and SPEA2~\cite{zitzler2001spea2}, exemplify this paradigm through repeated selection and variation. These methods have long been used as generic multi-objective optimizers. In modern protein design pipelines, similar iterative schemes are combined with deep generative models. RERD~\cite{uehara2025reward} repeatedly denoises protein sequences along a single optimization direction. Nevertheless,integrating iterative search with conditional diffusion under Pareto constraints remains underexplored for multi-objective optimization.

\section{Conclusions}

In this paper, we presented MP2D, a unified framework for multi-objective protein sequence generation and optimization that combines conditional diffusion with constrained MCTS and iterative refinement. By explicitly controlling Pareto candidate evolution and enabling iterative correction during inference, MP2D effectively addresses the challenges of optimizing multiple conflicting properties. Experiments on AMP and PB design tasks show that MP2D consistently achieves balanced improvements across multiple objectives and outperforms existing baselines without retraining generative models. In future work, MP2D can be extended to other protein design tasks. We believe this work provides a practical step toward scalable and reliable protein engineering.

\section*{Acknowledgements}

This research was partially supported by Fundamental and Interdisciplinary Disciplines Breakthrough Plan of the Ministry of Education of China No. JYB2025XDXM601, National Natural Science Foundation of China under Grant No. T2541004 and TIDRI under Grant No. KY052025003.

\bibliographystyle{named}
\bibliography{ijcai26}

\renewcommand{\thefigure}{S\arabic{figure}}
\renewcommand{\thetable}{S\arabic{table}}
\setcounter{figure}{0}
\setcounter{table}{0}

\clearpage
\noindent \textbf{\Large Appendix}
\appendix

\section{Implementation Details}

\subsection{Conditional Masked Diffusion Language Model (CMDLM)}

\subsubsection{Model Architecture}
As shown in Figure 1B, CMDLM is built upon an ESM-style transformer backbone with 30 layers, 20 attention heads, hidden size 640, and intermediate size 2560. The hidden dropout rate is set to 0.0, layer norm eps to 0.00001, max positional embedding to 1026 and GeLU activation functions. All models use standard amino-acid tokenization with a single [MASK] token from ESM-2~\cite{lin2022language}. The condition token is first projected with an added embedding layer and then summed with the embedded input after the embedding layer of ESM backbone. The condition token of general peptides, antimicrobial peptides and protein binders are set to 0, 1 and 2, respectively.

\subsubsection{Training Settings}
We trained CMDLM with the pipeline: We first pretrained CMDLM on general peptides and then performed conditional finetunes on antimicrobial peptides and protein binders. For pretraining on general peptides, we starts from the official weights of ESM2 150M, which are trained with the large-scale protein sequence database UniProtKB~\cite{uniprot2019uniprot}; then the model is pretrained unconditionally with full parameters on 2.6M peptide sequences from UniProt with sequence lengths between 2 and 50. Pretraining uses a masked diffusion objective with a linear noise schedule. For conditional funetunes on functional peptides, we leveraged LoRA on layer 19-29 with rank 16 and dropout 0.1. Classifier-free guidance is enabled by randomly dropping condition labels with probability 0.1 and cfg scale is set to 1.7.

We used a batch size of 64 and AdamW optimizer with beta (0.9,0.98) and weight decay of 0.01. We used a linear learning rate scheduler with 2000 warmup steps starting with 1e-7 and ending with 1e-3.

We trained CMDLM and ran the optimization program on only a single NVIDIA GeForce RTX 3090 GPU. All training tasks ended within 10,000 epoches and took within 120 hours.

\subsubsection{Diffusion Settings}
We used absorbing diffusion and the total number of diffusion steps is set to $T_{max}=10$. During generation, sequences are initialized as fully masked and iteratively denoised. For the noise schedule, we used linear schedule.

\subsection{Hyperparameter Settings}
The hyperparameter settings of CMCTD are summarized in Table S1 and the hyperparameter settings of global iterative refinement are summarized in Table S2.

\begin{table}[h]
\caption{Hyperparameter settings of CMCTD}
\centering
\resizebox{\linewidth}{!}{
\renewcommand\arraystretch{1.0}
\begin{tabular}{ccc}
\hline
Hyperparameter & Symbol & Value \\ \hline
noise adding level & T & $\text{Uniform}(1,10)$     \\
sequence length & L & 50     \\
$\#$ score functions for AMP & M & 4     \\
$\#$ score functions for PB & M & 5     \\
$\#$ direction vector & K & 64     \\
UCB exploration coeff & c & 0.1     \\
$\#$ child nodes & J & 30     \\
init angular threshold & $\Psi$ & $45^\circ$     \\
min angular threshold & $\Psi_{min}$ & $15^\circ$     \\
max angular threshold & $\Psi_{max}$ & $75^\circ$     \\
EMA smoothing coefficient & $\zeta$ & 0.5     \\
target rejection rate & $\eta$ & 0.3     \\ 
\hline
\end{tabular}
}
\label{tab:s1}
\end{table}

\begin{table}[h]
\caption{Hyperparameter settings of global iterative refinement}
\centering
\resizebox{\linewidth}{!}{
\renewcommand\arraystretch{1.0}
\begin{tabular}{ccc}
\hline
Hyperparameter & Symbol & Value \\ \hline
noise adding level & $T_{\tau,i}$ & $\text{Uniform}(0,10)$     \\
max iteration number & $\tau_{max}$ & 100     \\
$\#$ seed sequences & W & 100     \\
average weight for AMP & $\mu_{AMP}$ & [1,1,1,1]     \\
average weight for PB & $\mu_{PB}$ & [1,1,1,0.5,0.2]     \\
balancing weight & $\lambda$ & 1.0     \\
\hline
\end{tabular}
}
\label{tab:s2}
\end{table}

\section{MIC predictor}

\begin{figure}[h]
  \centering
  \includegraphics[width=0.6\linewidth]{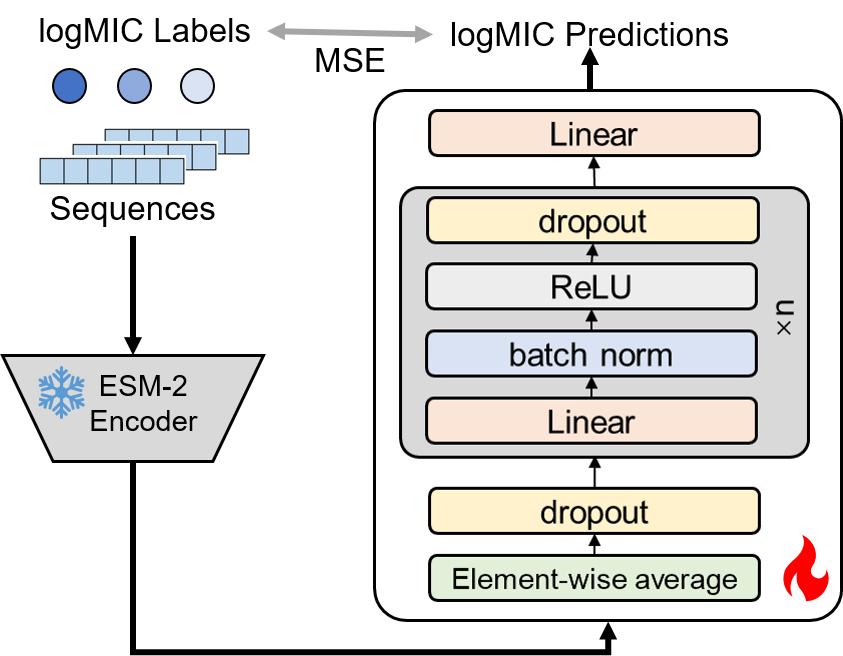}
  \caption{Architecture of the MIC regressor}
  \label{figs1}
\end{figure}

\subsection{Dataset}
We collected known AMP data with $\log$MIC values from the GRAMPA~\cite{plisson2022overcoming} and DBAASP~\cite{pirtskhalava2021dbaasp} datasets. We kept sequences with length less than 50 and without uncommon amino acids, and averaged the MIC values for difference speices per sequence. Finally we obtained 10,604 sequences with average MIC values; we splited it into training and validation set with the ratio of 9:1.

\subsection{Model Architecture}
As shown in Figure \ref{figs1}, the MIC predictor consists of a protein language model encoder and a modified multilayer perceptron (MLP). We use the ESM-2 650M encoder to transform input sequences into embeddings of size $L\times1280$, incorporating rich protein semantic information through the protein language model during this process. The MLP is composed of three layers, each consisting of a linear layer, a batch normalization layer, a ReLU activation layer, and a dropout layer. The first layer reduces the embedding dimension from 1280 to 1024, and each subsequent layer halves the input dimension. Finally, an output layer predicts the $\log$MIC with a single output dimension. Additionally, we include an element-wise average layer with dropout before the MLP. This layer compresses the $L\times1280$ embeddings into $1\times1280$ to integrate global sequence information, with the dropout rate set to 0.1.

\subsection{Training Settings}
We used a batch size 128, and trained for 200 epochs with a MSE loss. The optimizer we used is AdamW with beta value (0.9, 0.999), weight decay 0.01 and epsilon 0.000001. We used a constant learning rate scheduler with lr 0.0001. We evaluated the predictive performance of our regression model using Root Mean Squared Error (RMSE) and Coefficient of Determination ($R^2$). RMSE directly reflects the magnitude of the prediction error between the predicted and true values, while $R^2$ measures the correlation between the predicted and true values. A smaller RMSE and a larger $R^2$ indicate more accurate MIC predictions. The final model get $R^2=0.88891$ and $RMSE=0.2451$.

\section{Algorithm Details}

In this section, we describe the detailed algorithms of the fallback rules and constraint filtering strategies. Moreover, we describe the overall algorithm pipeline of Global Iterative Refinement and Constrained MCTS Diffusion.

\subsection{Fallback Rules}

Through the exponentially growing tree search, the generated sequences can explode rapidly. Especially when there are scalar objectives to be optimized, the Pareto optimal candidates will
directly adding these candidates can make Pareto front rapidly bloat and sparsify, making it difficult to search and distinguish better solutions and keep candidates with some properties that are too bad to accept. In this case, before updating the Pareto set and calculating the rewards, we propose to constrain based on the presupposed optimization directions. We first calculated the improvement vector $\Delta s_m$, indicating how much each individual objective improves for the action from $z_{t}$ to $z_{t-1,i}$
\begin{equation*}
    \Delta s_m(x_{t-1,i}) = s_m(x_{t-1,i}) - s_m(x_t)
\end{equation*}

Then we can get the angular differences between this improvement and the pre-defined optimization direction $\omega$ with
\begin{equation*}
    \gamma(z_{t-1,i}) = \frac{\Delta s(x_{t-1,i})\cdot \omega}{\left\| \Delta s(x_{t-1,i}) \right\|\left\| \omega \right\|}
\end{equation*}
The candidates that have angular differences less than a pre-define threshold $\Psi$ are kept. However, there might be no suitable candidates per step in practice, in which case we introduce a fallback rule: If there are no such candidates, we keep the ones $<\pi/2$. For the worst case that every $arccos(\gamma(z_{t-1,i}))\geq \pi/2$, indicating no possible action introduce optimization, we retain the current node $z_t$.
\begin{equation*}
    \text{child}'(z_t) = 
\begin{cases} 
\begin{split}
\{z_{t-1,i} | & arccos(\gamma(z_{t-1,i})) \leq \Psi\} \\& \text{if } \exists z_{t-1,i} \text{ s.t. } arccos(\gamma(z_{t-1,i})) \leq \Psi
\end{split}\\
\begin{split}
&\{z_{t-1,i}  | \Psi < arccos(\gamma(z_{t-1,i})) < \frac{\pi}{2}\} \\& \text{if } \nexists z_{t-1,i} \text{ s.t. } arccos(\gamma(z_{t-1,i})) \leq \Psi \\&\text{ and } \exists z_{t-1,i} \text{ s.t. } \Psi < arccos(\gamma(z_{t-1,i})) < \frac{\pi}{2}
\end{split}\\
\begin{split}
\{z_t\} & \text{ otherwise }
\end{split}\\
\end{cases}
\end{equation*}

Then we use the score vector to compute the rewards for each node in $\text{child}'(z_t)$

\subsection{Constraint Filtering}

A fixed filtering threshold $\Psi$ might not be optimal for different settings or study cases. In this case, we design a dynamic adaptation rule to adapt the filtering threshold $\Psi$ to a suitable value. We dynamically adjust it with a rejection rate $\xi_t = N(arccos(\gamma(z_{t-1,i}))>\Psi)/J$. In practice, we use an exponential moving average form
\begin{equation*}
    \overline{\xi}_t = \zeta \overline{\xi}_{t-h} + (1-\zeta)\xi_t
\end{equation*}
and we change $\Psi$ with
\begin{equation*}
    \Psi_{t+h} = 
\begin{cases} 
\Psi_{min} & \text{if } \Psi e^{\overline{\xi}_t-\eta}<\Psi_{min} \\
\Psi_{max} & \text{if } \Psi e^{\overline{\xi}_t-\eta}>\Psi_{max} \\
\Psi e^{\overline{\xi}_t-\eta} & \text{otherwise} \\
\end{cases}
\end{equation*}
where $\zeta\in [0,1)$ is a smoothing coefficient, $\overline{\xi}_0=\eta$ is the target rejection rate. This process dynamically increases the acceptable angular rage if too many candidates are rejected and vice versa.

\subsection{Global Iterative Refinement}

\begin{algorithm}
\caption{Global Iterative Refinement}
\begin{algorithmic}[1]
\STATE \textbf{Input:} Initial seed sequences $\{x_{0,i}\}_{i=1}^W$, Conditional diffusion model $p_\theta$, Property evaluators $\{s_m\}_{m=1}^M$, Optimization direction $\Omega$, max iteration $\tau_{max}$, max noise adding level $T_{max}$.
\STATE    Initialize current population $X_0 \leftarrow \{x_{0,i}\}_{i=1}^W$
\FOR{$0,1,\dots, \tau$ in $\tau_{max}$}
\STATE    Initialize next population $X_{\tau+1} \leftarrow \emptyset$
\FOR {each sequence $x_{\tau,i} \in X_\tau$}
\STATE    // Partial masking
\STATE    Sample noise level $T_{\tau,i} \sim \text{Uniform}(0, T_{max})$
\STATE    $z_{T_{\tau,i}} \leftarrow q(z_{T_{\tau,i}}|x_{\tau,i}, T_{\tau,i})$
\STATE    // Local optimization via CMCTD
\STATE    $\mathcal{P}^*_{\tau,i} \leftarrow \text{Constrained MCTS Diffusion}(z_{T_{\tau,i}}, p_\theta, {s_m}, \Omega)$
\STATE    // Candidate selection for next iteration
\IF{$|\mathcal{P}^*_{\tau,i}| = 1$}
\STATE    $x^*_{\tau,i} \leftarrow P^*_{\tau,i}$
\ELSE
\STATE    Compute $\Delta s(x)$ for all $x \in P^*_{\tau,i}$
\STATE    Compute $MPI(x)$ for all $x \in P^*_{\tau,i}$
\STATE    Sample $x^*_{\tau,i} \sim p(x) \propto exp(MPI(x))$ from $\mathcal{P}^*_{\tau,i}$
\ENDIF
\STATE    Add $x^*_{\tau,i}$ to $X_{\tau+1}$
\ENDFOR
\STATE    // Update population
\STATE    $X_\tau \leftarrow X_{\tau+1}$
\ENDFOR
\STATE \textbf{Output:} Final refined population $X_{\tau_{max}}$.
\end{algorithmic}
\end{algorithm}

\subsection{Constrained MCTS Diffusion}

\begin{algorithm}
\caption{Constrained MCTS Diffusion}
\begin{algorithmic}[2]
\STATE \textbf{Input:} Root partially masked sequence $z_T$, Conditional diffusion model $p_\theta$, Property evaluators $\{s_m\}_{m=1}^M$, Optimization direction $\omega$, Rollout budget $N_{rollout}$, Branching factor $J$, Angular threshold $\Psi$.
\STATE    // Initialization
\STATE    Pareto set $\mathcal{P}^* \leftarrow \emptyset$
\STATE    Initialize root node with state $z_T$
\STATE    Initialize visit count $N(\cdot) \leftarrow 0$ and reward $R(\cdot) \leftarrow 0$
\FOR {$1,\dots, n$ in $N_{rollout}$}
\STATE    // Selection
\STATE    $z \leftarrow z_T$
\WHILE{z is fully expanded and not terminal}
\STATE    $z \leftarrow argmax_{z'} UCB(z, z')$
\ENDWHILE
\STATE    // Expansion
\STATE    Generate J child nodes $\{z_{t-1,i}\}$ with $p_{\theta,i}(z_{t-1,i}^l|z_t^l)$
\STATE    // Simulation
\FOR {each child $z_{t-1,i}$}
\STATE    Fully denoise $z_{t-1,i}$ to obtain sequence $x_{t-1,i}$
\STATE    Evaluate property scores $s(x_{t-1,i})$
\STATE    Compute improvement vector $\Delta s(x_{t-1,i})$
\STATE    Compute angular similarity $\gamma(x_{t-1,i})$
\ENDFOR
\STATE    // Constraint Filtering
\STATE    Retain candidates satisfying $arccos(\gamma) \leq \Psi$
\IF{no candidate satisfies constraint}
\STATE    Retain candidates with $ \Psi < arccos(\gamma) < \pi/2$
\IF{still empty}
\STATE    Retain parent node $z$
\ENDIF
\ENDIF
\STATE    // Reward \& Pareto Update
\FOR {each retained candidate $x$}
\STATE    Compute reward vector $r(x)$ based on dominance over $\mathcal{P}^*$
\STATE    Update Pareto set $\mathcal{P}^*$ by adding non-dominated $x$
\STATE    Remove dominated elements from $\mathcal{P}^*$
\ENDFOR
\STATE    // Backpropagation
\FOR {each node $z'$ along path from expanded node to root}
\STATE    $R(z') \leftarrow R(z') + r(x)$
\STATE    $N(z') \leftarrow N(z') + 1$
\ENDFOR
\ENDFOR
\STATE \textbf{Output:} Pareto-optimal sequence set $\mathcal{P}^*$.
\end{algorithmic}
\end{algorithm}

\section{Visualization Analysis}

\subsection{Distributional Analysis of Optimization Effects}
Figure S2 visualizes the distributional changes of multiple properties before and after optimization for antimicrobial peptides (AMPs) and protein binders targeting 1B8Q and PPP5. Each violin plot compares the initial peptide distribution, task‑specific natural sequences, and sequences optimized by MP2D.

For AMP optimization, the optimized sequences exhibit a clear distributional shift toward higher antimicrobial activity and lower MIC values, while simultaneously improving non‑hemolysis and non‑toxicity. Notably, the optimized distributions become more concentrated with reduced variance compared to the initial peptide pool, indicating consistent improvements across candidates rather than isolated outliers. This suggests that MP2D effectively balances efficacy and safety objectives, which are known to be strongly conflicting in AMP design.

For protein binder optimization, similar patterns are observed for both targets. In the 1B8Q case, optimized binders show a pronounced upward shift in binding affinity, solubility, non‑fouling, and half‑life, while maintaining low hemolysis. Importantly, improvements are reflected across the entire distribution, rather than through extreme values alone, demonstrating stable multi‑objective trade‑offs.

For the more challenging PPP5 target, where characterized binders are scarce, MP2D still produces substantial distributional gains across all five properties. Although the initial peptide and PB baselines exhibit wide and skewed distributions, the optimized sequences consistently concentrate in favorable regions, particularly for affinity, solubility, and half‑life, highlighting the robustness of the optimization process under limited prior information.

Across all three tasks, MP2D does not merely shift one objective at the expense of others; instead, it reshapes the joint property distributions toward balanced, high‑quality regions. These results visually corroborate the quantitative findings and demonstrate MP2D’s ability to perform robust multi‑objective optimization by improving both central tendencies and distributional consistency.

\begin{figure}[h]
  \centering
  \includegraphics[width=\linewidth]{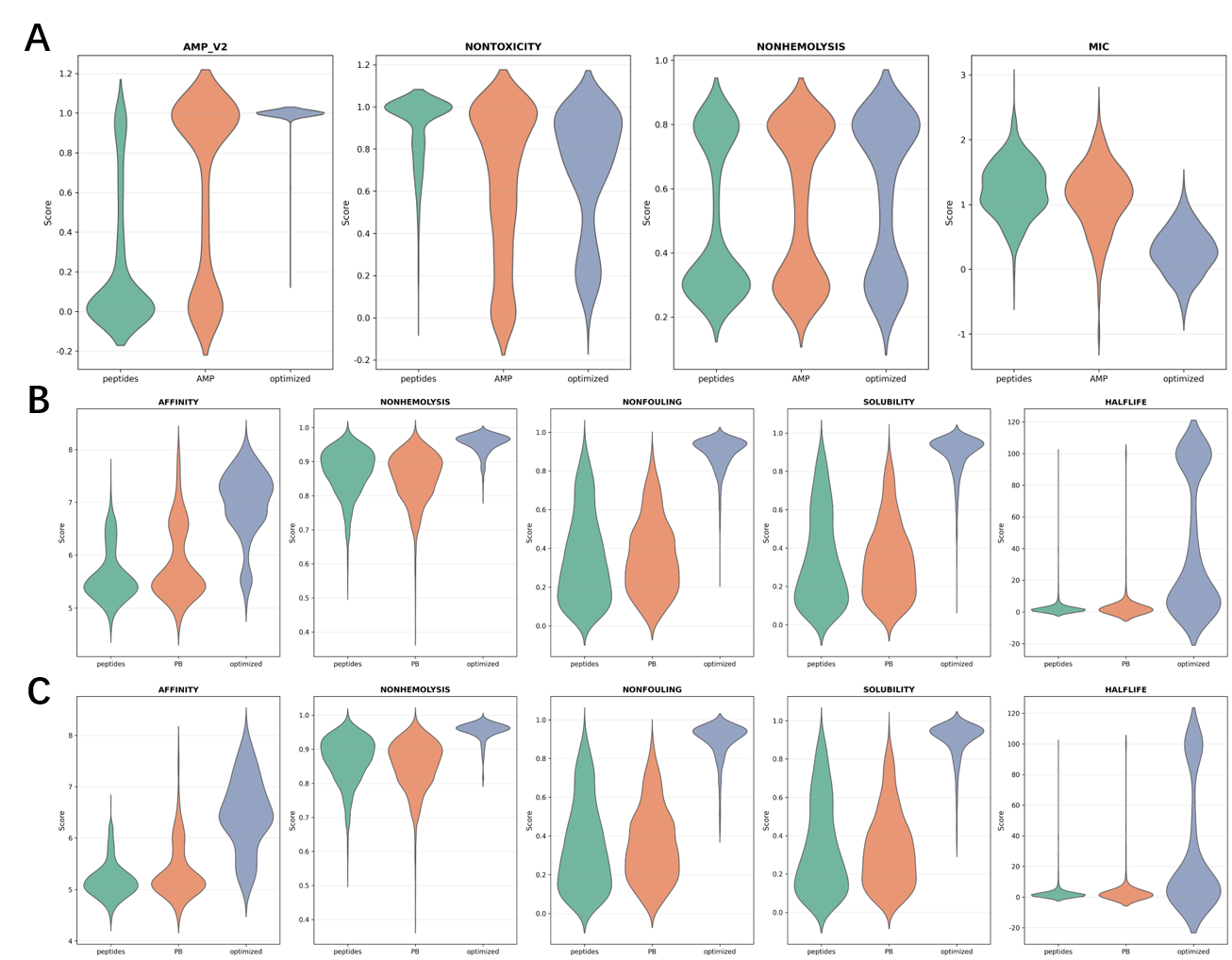}
  \caption{Visualization of the distributional changes of multiple properties before and after optimization for (A) AMPs; (B) protein binders targeting 1B8Q; and (C) protein binders targeting PPP5}
  \label{figs2}
\end{figure}

\subsection{Structural Case Studies of Designed Protein Binders}

To further examine the structural plausibility of the designed protein binders, we selected representative optimized sequences for targets 1B8Q and PPP5 and predicted their complex structures using AlphaFold‑Multimer. Figure S3 visualizes the predicted binding modes, where the target proteins are shown in surface representation and the designed binders are highlighted.

For both targets, the predicted complexes exhibit interface predicted TM‑scores (ipTM) around 0.7, indicating confident and consistent interface formation between the designed binders and their respective targets. Such ipTM values are commonly considered indicative of reliable interfacial geometry in protein–protein interaction prediction, suggesting that the optimized sequences are structurally compatible with the target binding sites.

In the 1B8Q case, the designed binder adopts a compact conformation that fits well into a surface groove of the target protein, forming an extended interface. This structural arrangement is consistent with its high predicted affinity and favorable solubility and half‑life, while maintaining low hemolysis.

For the more challenging PPP5 target, despite the lack of characterized natural binders, the optimized sequence also forms a stable and well‑defined interface, with the binder spanning multiple contact regions on the protein surface. The predicted complex aligns with the strong improvements observed in affinity, solubility, non‑fouling, and half‑life.

These structural case studies provide qualitative evidence that MP2D not only improves scalar property scores but also yields binders with physically plausible binding modes. Combined with the distributional and quantitative evaluations, the results suggest that MP2D can generate protein binders that are both functionally optimized and structurally coherent.

\begin{figure}[h]
  \centering
  \includegraphics[width=\linewidth]{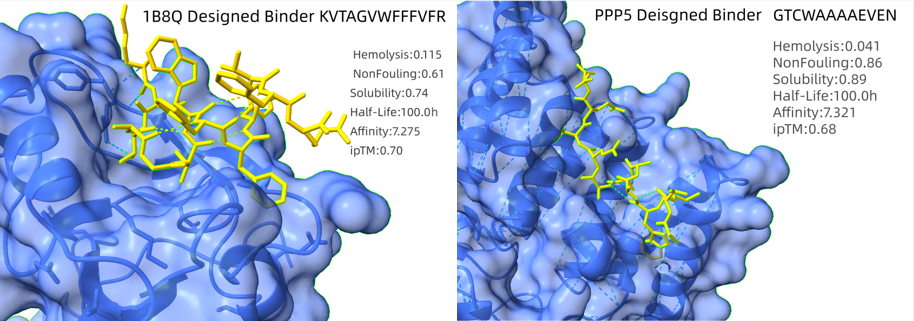}
  \caption{Structures of complex between designed binders with (left) 1B8Q and (right) PPP5.}
  \label{figs3}
\end{figure}

\section{Baselines}
In this appendix, we briefly describe the baseline methods used in our experiments and explain their relevance to protein sequence generation and multi-objective optimization.

\subsection{Baselines for Conditional Diffusion Backbone Evaluation}
\textbf{ProtGPT2}~\cite{ferruz2022protgpt2}  is an autoregressive protein language model adapted from the GPT-2 architecture and trained on large-scale protein sequence data. It generates protein sequences token by token and captures long-range sequential dependencies. We include ProtGPT2 as a representative autoregressive baseline to compare diffusion-based generation against classical likelihood-based protein language models.

\textbf{ProteinGAN}~\cite{repecka2021expanding} is a generative adversarial network designed for de novo protein sequence generation. It learns an implicit distribution over protein sequences through adversarial training and is capable of producing diverse sequences resembling natural proteins. ProteinGAN is selected as a representative GAN-based approach, providing a contrast to both autoregressive and diffusion-based generative paradigms.

\textbf{EvoDiff}~\cite{alamdari2023protein} is a discrete diffusion model specifically developed for protein sequence generation. By iteratively denoising corrupted sequences, it enables gradual refinement rather than single-pass generation. EvoDiff represents the state of the art in diffusion-based protein generation and serves as the most directly comparable baseline to CMDLM.

All backbone baselines are configured with comparable parameter sizes and standard character-level tokenization following a simplified version of the benchmark protocol of~\cite{meshchaninov2024diffusion}, ensuring a fair comparison of generative quality. We note that more recent protein diffusion models, such as DPLM~\cite{wang2024diffusion}, have been proposed. Our framework is flexible and open to be replaced with new diffusion models as backbones, which might bring further improvements. However, our focus is on establishing a strong and representative diffusion backbone under a standardized benchmark, which suffices for validating the effectiveness of the proposed optimization framework.

\subsection{Baselines for Protein Binder Optimization}

\textbf{MOG-DFM}~\cite{chen2025multi} is a recent discrete flow matching–based method proposed specifically for protein binder optimization. It guides sequence generation by adaptively computing probability paths under multiple objectives. We include MOG-DFM as a strong generative optimization baseline that combines deep generative modeling with multi-objective design.

Moreover, MOG-DFM provides a comprehensive and complete benchmark setup. We directly follow this benchmark to evaluate our method. These concluded baselines collectively represent classical evolutionary and swarm-based methods for protein binder design:

\textbf{NSGA-III}~\cite{ishibuchi2016performance} is a many-objective evolutionary algorithm based on non-dominated sorting and reference directions to maintain solution diversity. It is widely used for optimizing problems with more than three objectives. We include NSGA-III as a standard Pareto-based evolutionary baseline for many-objective protein binder optimization.

\textbf{SMS-EMOA}~\cite{beume2007sms} is an indicator-based multi-objective evolutionary algorithm that directly optimizes hypervolume contribution. By explicitly encouraging both convergence and diversity, it serves as a strong baseline for Pareto-front optimization. SMS-EMOA is selected to represent hypervolume-driven evolutionary optimization.

\textbf{SPEA2}~\cite{zitzler2001spea2} is an elitist multi-objective evolutionary algorithm that combines dominance strength and density estimation to guide selection. It maintains an external archive of non-dominated solutions. SPEA2 is included as a classical and widely adopted Pareto-based optimizer.

\textbf{MOPSO}~\cite{coello2002mopso} extends particle swarm optimization to the multi-objective setting using a swarm of particles and an archive of non-dominated solutions. It represents the class of swarm-intelligence-based optimizers and provides a complementary baseline to evolutionary algorithms.

\subsection{Baselines for Antimicrobial Peptide Optimization}

To the best of our knowledge, no standardized benchmark exists for multi-objective AMP optimization. Therefore, we select the following baselines to cover diverse modeling strategies, including conditional generation, discriminator-based optimization, online retraining, and multimodal learning.

\textbf{Multi-CGAN}~\cite{yu2023multi} is a conditional GAN-based model for multi-objective antimicrobial peptide design. It incorporates multiple property labels during training and generates peptides conditioned on desired label combinations. We include Multi-CGAN as a representative label-conditioned generative baseline.

\textbf{HMAMP}~\cite{wang2025hmamp} employs multiple discriminators, each corresponding to a specific AMP property, and trains the generator to satisfy all discriminators simultaneously. This model exemplifies discriminator-driven multi-objective optimization during training and is included as a strong GAN-based AMP design baseline.

\textbf{MPOGAN}~\cite{liu2025multi} performs multi-objective optimization by iteratively retraining the generator on samples filtered by multiple property classifiers. This online retraining strategy allows gradual improvement of target properties but requires repeated model updates. MPOGAN is included to represent retraining-based multi-objective optimization.

\textbf{MoFormer}~\cite{wang2024moformer} is a multimodal fusion model that integrates sequence representations with auxiliary property information for AMP design. It represents transformer-based multimodal approaches to antimicrobial peptide optimization and provides a complementary baseline to GAN-based methods.

We select these baselines because they represent the most recent and competitive methods specifically designed for antimicrobial peptide optimization under multiple properties. In particular, they are among the few approaches that can handle more than three AMP-specific objectives simultaneously. General-purpose multi-objective optimizers or protein binder–oriented methods are not included, as they rely on property predictors and optimization assumptions that are not directly applicable to AMP design, where objectives, data distributions, and evaluation models differ substantially.

\section{Metrics}
In this appendix, we describe the evaluation metrics used in our experiments and explain the rationale for selecting each metric in different evaluation settings.

\subsection{Foundation Model Evaluation}
To assess the quality of the conditional masked diffusion language model (CMDLM), we evaluate generation performance from three complementary perspectives: sequence plausibility, structural foldability, and distributional similarity.

\textbf{ESM‑2 Perplexity}~\cite{lin2022language} measures how well generated sequences align with the statistical patterns learned by a large pretrained protein language model. Lower perplexity indicates higher sequence plausibility and consistency with natural protein sequences. We adopt this metric as a standard proxy for sequence-level realism in protein generation. Here, we calculate ESM-2 pppl with ESM-2 35M by masking each amino acid of the protein sequence and predicting it considering all the other amino acids in the sequence, and calculating the value with the equation
\begin{equation*}
\begin{split}
    & \text{ESM-2 pseudoperplexity} = \\& \exp{( -\frac{1}{|x|} \sum_{i=1}^{|x|} \log p(x_i \mid x_{j \neq i}, \theta_{\text{ESM-2}}) )}
\end{split}
\end{equation*}

\textbf{Predicted Local Distance Difference Test (pLDDT)}~\cite{jumper2021highly} scores are obtained from structure prediction models and reflect the confidence of predicted protein structures. In this study we utilize OmegaFold~\cite{wu2022high} to compute the pLDDT value. Higher pLDDT values indicate better structural foldability and stability. We use pLDDT to assess whether generated sequences are likely to form well-defined three-dimensional structures.

Frechet ProtT5 Distance (FPD).
FPD measures the distance between embedding distributions of generated sequences and real protein sequences using ProtT5 representations. Given two samples $X_1 \sim \mathcal{N}(\mu_1, \Sigma_1)$ and $X_2 \sim \mathcal{N}(\mu_2, \Sigma_2)$, the FPD can be calculated as 
\begin{equation*}
    FPD = \lVert \mu_1 - \mu_2 \rVert^2 + \text{tr}(\Sigma_1 + \Sigma_2 - 2\sqrt{\Sigma_1 \Sigma_2})
\end{equation*}
This metric captures distributional similarity beyond individual sequences and is commonly used to evaluate whether generated samples resemble the target protein class as a whole.

Together, these metrics provide a comprehensive evaluation of generative quality at the sequence, structure, and distributional levels.

\subsection{Protein Binder Optimization}

For protein binder optimization, we follow the benchmark of MOG-DFM to evaluate five therapeutically relevant properties that are commonly considered in peptide and protein drug development:
\begin{itemize}
\item Hemolysis: measuring red blood cell toxicity;
\item Non‑fouling: indicating resistance to nonspecific interactions;
\item Solubility: reflecting formulation and bioavailability;
\item Half‑life: representing in vivo stability;
\item Binding affinity: quantifying target‑specific interaction strength.
\end{itemize}
All properties are evaluated using publicly available pretrained classifiers following the benchmark protocol of~\cite{liu2025multi}. These predictors are widely used in protein binder design and provide a standardized and reproducible evaluation framework. Using the same evaluators across methods ensures fair comparison under consistent objective definitions.

\subsection{Antimicrobial Peptide Optimization}
Antimicrobial peptide (AMP) optimization focuses on properties that are highly specific to antimicrobial activity and safety. We evaluate four key AMP‑relevant properties using established public predictors:
\begin{itemize}
\item Antimicrobial probability ($P_{amp}$): evaluated using AMP‑Scanner~\cite{veltri2018deep}, which predicts the likelihood that a peptide exhibits antimicrobial activity;
\item Hemolysis: evaluated using HemoPI2~\cite{rathore2025prediction}, to assess red blood cell toxicity;
\item Toxicity: evaluated using ToxinPred3~\cite{rathore2024toxinpred}, to estimate general cytotoxic risk.
\item Minimum inhibitory concentration (MIC): evaluated using a regression model trained on experimentally measured MIC data. MIC directly reflects antimicrobial potency and provides a quantitative complement to binary activity prediction.. 
\end{itemize}

These metrics are selected because they represent the most widely adopted and biologically meaningful evaluation criteria for AMP design, and together capture both antimicrobial efficacy and safety considerations.

\end{document}